\documentclass[10pt,letterpaper]{article}
\usepackage[top=0.85in,left=2.75in,footskip=0.75in]{geometry}

\usepackage{amsmath,amssymb}

\usepackage{changepage}

\usepackage[utf8x]{inputenc}

\usepackage{textcomp,marvosym}

\usepackage{cite}

\usepackage{nameref,hyperref}

\usepackage[right]{lineno}

\usepackage{microtype}
\DisableLigatures[f]{encoding = *, family = * }

\usepackage[table]{xcolor}

\usepackage{array}

\newcolumntype{+}{!{\vrule width 2pt}}

\newlength\savedwidth

\usepackage{setspace} 
\raggedright
\usepackage[aboveskip=1pt,labelfont=bf,labelsep=period,justification=raggedright,singlelinecheck=off]{caption}

\makeatletter
\renewcommand{\@biblabel}[1]{\quad#1.}
\makeatother

\date{}

\usepackage{lastpage,fancyhdr,graphicx}
\usepackage{epstopdf}

\begin{document}
\vspace*{0.2in}

\begin{flushleft}
{\Large
\textbf\newline{Chemotactic drift speed for bacterial motility pattern with two alternating turning events} 
}
\newline
\\
Evgeniya V. Pankratova\textsuperscript{1*},
Alena I. Kalyakulina\textsuperscript{1},
Mikhail I. Krivonosov\textsuperscript{1},
Sergei V. Denisov\textsuperscript{1, 2},
Katja M. Taute\textsuperscript{3},
Vasily Yu. Zaburdaev\textsuperscript{4, 5}.
\\
\bigskip
\textbf{1} Institute of Information Technologies, Mathematics and Mechanics, Lobachevsky State University, Nizhniy Novgorod, Russia
\\
\textbf{2} Department of Theoretical Physics, University of Augsburg, Germany
\\
\textbf{3} Rowland Institute at Harvard, Harvard University, Cambridge, USA
\\
\textbf{4} Max Planck Institute for the Physics of Complex Systems, Dresden, Germany
\\
\textbf{5} Institute of Supercomputing Technologies, Lobachevsky State University, Nizhniy Novgorod, Russia
\\
\bigskip

*E-mail: evgenia.pankratova@itmm.unn.ru

\end{flushleft}
\section*{Abstract}
Bacterial chemotaxis is one of the most extensively studied adaptive responses in cells. Many bacteria are able to bias their apparently random motion to produce a drift in the direction of the increasing chemoattractant concentration. It has been recognized that the particular motility pattern employed by moving bacteria has a direct impact on the efficiency of chemotaxis. The linear theory of chemotaxis pioneered by de Gennes allows for calculation of the drift velocity in small gradients for bacteria with basic motility patterns. However, recent experimental data on several bacterial species highlighted the motility pattern where the almost straight runs of cells are interspersed with turning events leading to the reorientation of the cell swimming directions with two distinct angles following in strictly alternating order. In this manuscript we generalize the linear theory of chemotaxis to calculate the chemotactic drift speed for the motility pattern of bacteria with two turning angles. By using the experimental data on motility parameters of {\em V. alginolyticus } bacteria we can use our theory to relate the efficiency of chemotaxis and the size of bacterial cell body. The results of this work can have a straightforward extension to address most general motility patterns with alternating angles, speeds and durations of runs.


\section*{Introduction}
Bacteria are the most numerous living organisms \cite{Whitman1998}. A variety of shapes, sizes and ways of movement enable them to adapt to different environmental conditions \cite{Kearns2010, OToole2000}. One of the most common forms of bacteria existence are biofilms, which are multicellular colonies with a complex spatial and metabolic structure forming at interfaces. Ability of individual cells to move and sense environmental signals are crucial for cell aggregation and biofilm formation \cite{Stoodley2002}. Bacteria can move on solid surfaces or swim in liquid media \cite{Jarrell2008}. Although these movements often look like a random motion, bacteria can bias this random motion to move in a certain direction on average. One well-known example of such directed motion is chemotaxis -- the ability to alter motility in response to gradients of chemicals \cite{Eisenbach2004}. Bacteria in homogeneous environments often exhibit very particular motility patterns, which can greatly affect their ability to perform chemotaxis \cite{Taktikos2013}. 

One of the most studied motility pattern is \textquotedblleft{}run-and-tumble\textquotedblright{} of \textit{Escherichia coli} \cite{Berg1972}. \textit{E. coli} uses multiple rotating flagella to swim. When all flagella rotate counterclockwise, they form a bundle that drives the bacterium in an approximately straight trajectory of a \textquotedblleft{}run\textquotedblright{}. When one or more flagella begin to rotate clockwise, the bundle breaks up leading to a change of the swimming direction, known as \textquotedblleft{}tumble\textquotedblright{} \cite{Turner2000}. For \textit{E. coli} the angle between the next and the previous directions is randomly distributed with a mean of approximately $62^{\circ}$ \cite{Berg1972}. Many bacteria species, in particular those with a single flagellum, completely reverse the direction of their motion after switching of flagellum rotation, thus leading to so called “run-reverse” motility pattern \cite{Theves2013, Johansen2002}. 

Importantly, bacteria are able to alternate their motility pattern in response to gradients of certain signaling chemicals. Swimming cells sense the concentration of the signal and extend the duration of the runs, when moving in the direction of the chemical gradient. The chemotactic response of the cells is usually quantified by the average drift velocity in the direction of the gradient. After the key result of de Gennes, who proposed the so called linear theory of chemotaxis, the chemotactic drift speed was calculated for some basic motility patterns of bacteria \cite{Taktikos2013, Locsei2007, deGennes2004}.

Recently advances in bacteria tracking and a careful analysis of bacterial trajectories led to the discovery of more complex motility patterns. For example, a bacterium \textit{V. alginolyticus} exhibits a \textquotedblleft{}run-reverse-flick\textquotedblright{} pattern with two alternating average turning angles \cite{Xie2011, Son2013, Berg2013}. Based on the ensemble measurements it was previously believed that these angles were $180$ and $90$ degrees \cite{Xie2011}. However, new data shows, that while the reversal is indeed universal for all cells, the second turning angle is cell-size dependent \cite{Son2013} and varies significantly between individuals \cite{Taute2015}. Importantly previous analytical results for the drift speed of \textit{V. alginolyticus} swimming pattern were obtained under assumption of the second turning (flick) angle of $90^{\circ}$, which dramatically simplifies calculations \cite{Taktikos2013}.

In this manuscript, we provide an analytical calculation of the drift speed of chemotactic bacteria moving in a pattern with two alternating arbitrary turning angles. It is thus, to the best of our knowledge, the most general to date extension of the de Gennes result that can be applied to a broad class of bacterial motility patterns. Furthermore it allows us to predict the cell-to-cell variability in the drift speed of \textit{V. alginolyticus} based on published experimental data on the cell-size motility dependence \cite{Taute2015}.

In the following Section \hyperref[S2]{II} we outline the derivation of the main result and in Section \hyperref[S3]{III} combine it with experimental data on \textit{V. alginolyticus} swimming pattern. Section \hyperref[S4]{IV} is reserved for discussions.  

\section*{Chemotactic drift speed calculation for a swimming pattern with two alternating turning events}
\label{S2}

Various bacteria utilize distinct swimming patterns to navigate their environment. Some of these patterns can be considered as two-step processes (\textquotedblleft{}run-and-tumble\textquotedblright{} pattern of \emph{E. coli}, for instance), or as four-step processes (as \textquotedblleft{}run-reverse-run-flick\textquotedblright{}, or \textquotedblleft{}run-reverse-flick\textquotedblright{} for short, pattern of \emph{V. alginolyticus})~\cite{Xie2011}. In the \textquotedblleft{}run-reverse-flick\textquotedblright{} pattern, a cell swims forward for some time interval and it then backtracks by reversing the direction of the flagellar motor rotation. However, upon resuming forward
swimming, the flagellar hook experiences mechanical instability and flicks, causing the cell body to reorient and choose a new swimming direction \cite{Son2016}. Recent experiments show that the average flick angle is cell-size dependent with larger cells having larger flick angles \cite{Taute2015} (meaning that larger cells stabilize the flick by counteracting viscous drag force acting on the cell body and their turning angle is closer to reversal). The general theoretical approach presented in this work allows for an analytical treatment of the corresponding chemotactic strategy through the universal de Gennes formalism \cite{deGennes2004}. We now formulate the model for the bacteria motion with two alternating turning events.

\textbf{The pattern of $\alpha-\beta$ bacteria motion}

Let the swimming pattern of bacteria consist of 4 phases: \textquotedblleft{}Run 1\textquotedblright{} -- movement along a certain direction, \textquotedblleft{}$\alpha$-rotation\textquotedblright{} -- changing the direction of subsequent movement by a random angle $\Delta \varphi_1$ with a corresponding average cosine value denoted by $\alpha$, $\alpha \equiv \langle\cos \Delta\varphi_1 \rangle$, \textquotedblleft{}Run 2\textquotedblright{} -- movement along the new direction, \textquotedblleft{}$\beta$-rotation\textquotedblright{} -- changing the direction of the subsequent movement by another random turning angle $\Delta\varphi_2$ with the average cosine denoted by $\beta$, $\beta \equiv \langle\cos \Delta\varphi_1 \rangle$. The speed of the bacterial movement between two subsequent rotations is considered to be constant and the same for both runs. Interestingly, there are reported cases when the forward and backward swimming speeds are different \cite{Theves2013, Theves2015}. Different speeds can also be included in the model, however here we keep them the same to focus on the effect of two alternating angles. 

Despite the fact that the times at which the turning events occur are stochastic, the sequence at which the types of turning events follow each other is fixed. Note that, the pattern of $\alpha-\beta$ bacteria motion considered in this paper has been observed not only in \textit{V. alginolyticus} but also in other marine bacteria species with a single flagellum \cite{Xie2011, Son2013, Liu2014}. Since it was shown by Altindal et al \cite{Altindal2011} that in the absence of chemical gradient \textit{V.alginoliticus} demonstrates equal mean values for durations of runs after both turning events and flicks (with $\tau \approx 0.3 $s), in our model we assume the running time after both types of reorientation events to be exponentially distributed with the same mean $\tau_{run}$. We should note that the run time distribution measured in experiments often deviates from exponential for short run times. In this respect the exponential distribution is a simplifying assumption, which is, however, crucial for analytical feasibility of the following calculations \cite{footnote1}. As with two alternating speeds, two different run time distributions can be included without conceptual difficulties but at a cost of lengthier expressions. Durations of both turning events are one order of magnitude shorter than runs and thus usually neglected in the analysis, see however \cite{Kafri2008}. In our model we treat turnings as instantaneous.

During every run the swimming direction of the cell fluctuates due to the thermal noise in the fluid and possibly due to active processes in the flagellar motor. This effect can be characterized by means of rotational diffusion with a constant $D_r$. The value of $D_r$ can be measured experimentally and in general is in agreement with the estimate of Brownian rotational diffusion of a passive particle of the size of the cell \cite{Berg1972}. In general the effect of rotational diffusion is not small; a typical deviation of the swimming direction during a single run can be of the order of $30^{\circ}$ \cite{Berg1972}.

\begin{figure}[ht!!!]
	\begin{center}
		\includegraphics[width=100mm]{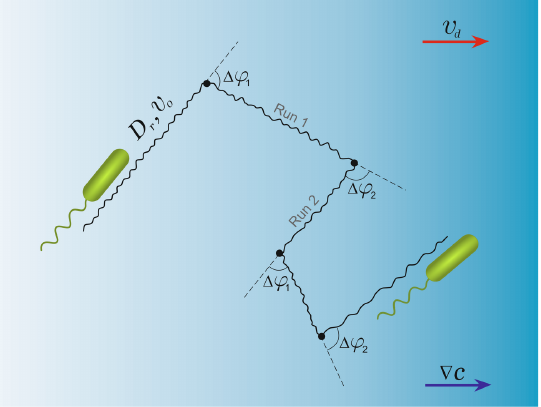}
				\caption{\textbf{Motility pattern with two turning events.} \hspace{\textwidth}During each run, the speed of the cell $v_0$ is constant. Motion is nearly straight and is affected by the rotational diffusion $D_r$. The cell changes the direction of its motion during turning events (black dots), where turning angles $\Delta \varphi_{1, 2}$ are allowed to have two different probability distributions. Importantly, the cell strictly alternates the two types of turning events. When swimming in the gradient of signaling chemicals $\nabla c$, the cell can bias its motion and respond with a drift speed $v_d$ in the direction of the gradient, which we want to calculate.}
		\label{fig1}
	\end{center}
\end{figure}

A sketch of the trajectory of $\alpha-\beta$ pattern is provided in Fig.~\ref{fig1}. In the absence of signaling chemicals it is a trajectory of a random walk which can be characterized by a velocity correlation function and the effective long time diffusion constant (see \cite{Taktikos2013} for the corresponding calculations). 

\subsection*{Effect of chemotaxis}

In the presence of a chemical gradient bacteria are able to direct their overall random motion towards the
attractant. Bacteria possess a chemo-sensory system allowing for temporal integration of the external chemical cues and a delayed response which biases the rotational direction of flagellar motors. When the cell climbs up the gradient it can extend durations of the run phases. The biochemical structure of the chemotaxis response is well understood at least for \textit{E.coli} and many of its features are known experimentally. Here we use the linear model of chemotaxis proposed by de Gennes \cite{deGennes2004}. This model postulates that the turning frequency of the bacterium is influenced by the 
experienced concentration of the chemical $c(t)$ in a following simple form: 

\begin{equation}
\label{f1}
\lambda(t) = \lambda_0 \left( 1 - \int_{-\infty}^{t} R(t-\tau) c(\tau) d\tau \right) ,
\end{equation}
where $\lambda_0 = 1/\tau_{run}$ and $R(t)$ is the internal response of the cell, which was measured experimentally for \textit{E. coli} \cite{Block1983}. A typical functional form of the response kernel following from experiments and some analytical arguments on the optimality of the response \cite{Celani2009} is:
\begin{equation}
\label{f2}
R(T) = W \lambda_0 e^{-\lambda_0 T} \left[ 1-\frac{\lambda_0 T}{2} - \left( \frac{\lambda_0 T}{2} \right)^2 \right],
\end{equation}
where $W$ is a constant characterizing the strength of the response and has the dimension of volume. The central feature of this response kernel is the property

\begin{equation}
\label{f3}
\int_{-\infty}^{\infty} R(\tau) d\tau = 0,
\end{equation}
recognized as the ability of cells to adapt to background concentration of chemicals and sense small gradients even on the high background levels of the signal. The vanishing of the integral means that any background concentration, which is spatially homogeneous, will not affect the tumbling frequency, as seen from Eq.(\ref{f1}). Thus, by adopting the response with a zero integral the cell gains the ability to detect small concentration variations independent of the overall constant background concentration of the signaling molecule. Importantly a similar response function was experimentally measured for \textit{V. alginolyticus} bacteria \cite{Xie2015}. Although the response for backward and forward motion might differ by a numerical factor of the order of $2$ \cite{XieLu2015}, here for simplicity of calculations we assume the same response for both swimming directions. To quantify the effectiveness of chemotaxis we use the so-called chemotactic drift velocity $v_d$ in the concentration of the chemoattractant with a small linear gradient pointing along $Oz$ axis:

\begin{equation}
\label{f4}
c(z) = |\nabla c| z,
\end{equation}
where $|\nabla c| = \text{const}$ (constant gradient). Without loss of generality we consider a random walk whose first reorientational event at $t=0$ is an \textquotedblleft{}$\alpha$-rotation\textquotedblright{}, and, correspondingly, with the turning of type $\beta$ at $t=t_{\beta}>0$. Then we can determine the drift velocity as the sum of average displacements of a run $\langle \overline{z}_{\beta} \rangle$ and a subsequent run $\langle \overline{z}_{\alpha} \rangle$, divided by mean duration of two runs $2\tau_{run}$:

\begin{equation}
\label{f5}
v_d = \frac{\langle \overline{z}_{\beta} \rangle + \langle \overline{z}_{\alpha} \rangle}{2 \tau_{run}}.
\end{equation}
For $\langle \overline{z}_{\alpha} \rangle$ and $\langle \overline{z}_{\beta} \rangle$ we calculate the expectation over all possible paths taking into account that the position of the bacterium $z(t)$ at a time $t$ on a particular path is random:

\begin{equation}
\label{f6}
\langle \overline{z}_{\beta} \rangle = \int_{0}^{\infty} \langle  z(t) p(t) \rangle dt,
\end{equation}
where $p(t)$ is the probability density function for the time corresponding to the run termination event at $t=t_{\beta}$ on a particular path. Obviously the drift velocity of the bacterial population with $\alpha=\beta$ becomes $v_{d} = \langle \overline{z} \rangle / \tau_{run}$.

As was shown by de Gennes, we can first analyze the drift velocity $v_{\delta}$ for a simplified response kernel: 

\begin{equation}
\label{f7}
R(t) =A\delta(t-T)
\end{equation}
with a delay time $T$ and a strength $A$ \cite{Locsei2007}, and then obtain the desired drift speed with the full kernel Eq.(\ref{f2}) by a simple integration:

\begin{equation}
\label{f8}
v_{d} = \int_{0}^{\infty} R(T) \frac{v_\delta (t)}{A} dT,
\end{equation}
where $v_\delta$ is the chemotactic drift speed for the delta-response Eq.(\ref{f7}). While we give full details of the corresponding derivation in \hyperref[S1_Appendix]{S1 Appendix}, here we want to outline the conceptual steps of this calculation. 

The derivation of de Gennes \cite{deGennes2004} relies on the exact answer for the mean run time in case of time dependent turning rate. This is expanded up to the linear order in the gradient of concentration. Finally this gradient can be related to the position of the particle. In both de Gennes' calculations \cite{deGennes2004} and previous results on \textit{V. alginolyticus} presence of a $90^{\circ}$ turning angle \cite{Taktikos2013} significantly simplified calculations as a turn by $90^{\circ}$ completely randomizes new direction and thus erases memory. For a two-step process of \textit{E.coli} with an arbitrary tumble angle and therefore with the non-disappearing memory the problem was solved by Locsei \cite{Locsei2007}. Our goal is to extend these results to the 4-step pattern. In this case, assuming that the cell has been swimming in the chemical gradient for rather long time, we can estimate the drift velocity via the Eq.(\ref{f5}). However, in general (for two alternating turning events with arbitrary angles) the integrals for $\langle \overline{z}_{\beta}\rangle$ and $\langle \overline{z}_{\alpha}\rangle$ become dependent on each other. This peculiarity is one of the central technical difficulties that should be taken into account. Another important point is the integrand transformation in the Eq.(\ref{f6}) alowing to reduce it to an integrable form. Similarly to \cite{Taktikos2013} and \cite{Locsei2007}, to obtain the general expression for the emerged velocity autocorrelation function $\langle v_z(t)v_z(t') \rangle$ being valid for any $t$ and $t'$, we decomposed it for separate intervals of motion. Taking into account the alternating feature of the considered random walk process, the multipliers in this decomposition can be represented in terms of both $\alpha$- and $\beta$-type turnings. Integration of the obtained functions gives the expressions for $\langle \overline{z}_{\beta}\rangle$ and $\langle \overline{z}_{\alpha}\rangle$ whose combination according to Eq.(\ref{f5}) leads to the following formal analytical result for the drift velocity in the case of delta-response:

\begin{equation}
\label{f9}
v_\delta = \frac{v_0^2}{3} \lambda_0 A|\nabla c| \left[ k_\delta \cosh{\left(\sqrt{\alpha}\sqrt{\beta}\lambda_0 T\right)} + m_\delta \sinh{\left(\sqrt{\alpha}\sqrt{\beta}\lambda_0 T\right)} + n_\delta \right]
\end{equation}
with the coefficients $k_\delta$, $m_\delta$, $n_\delta$: 

\begin{equation}
\label{f10}
\begin{array}{ll}
\displaystyle{k_\delta = \frac{\lambda_0 e^{-\left(2D_r+\lambda_0\right)T} \left( 4D_r^2\left(2-s_{\alpha\beta}\right) - 8D_r d_{\alpha\beta}\lambda_0 - \left(2+s_{\alpha\beta}\right)d_{\alpha\beta}\lambda_0^2 \right)}{2\left(4D_r^2+4D_r\lambda_0-d_{\alpha\beta}\lambda_0^2\right)^2}}, \\ 

\displaystyle{m_\delta = \frac{\lambda_0 e^{-\left(2D_r+\lambda_0\right)T} \left( 4D_r^2\left(s_{\alpha\beta}-2\alpha\beta\right) - 4D_rd_{\alpha\beta}\lambda_0 s_{\alpha\beta} - \lambda_0^2d_{\alpha\beta}\left(s_{\alpha\beta}+2\alpha\beta\right) \right)}{2\left(4D_r^2+4D_r\lambda_0-d_{\alpha\beta}\lambda_0^2\right)^2 \sqrt{\alpha}\sqrt{\beta}}}, \\

\displaystyle{n_\delta = \frac{\lambda_0^2 \left( 4D_r^2\left(-s_{\alpha\beta}+2\alpha\beta\right) + 4D_r\lambda_0d_{\alpha\beta}s_{\alpha\beta} + \lambda_0^2d_{\alpha\beta}\left(s_{\alpha\beta}+2\alpha\beta\right) \right)} {2\left(2D_r+\lambda_0\right)\left(4D_r^2+4D_r\lambda_0-d_{\alpha\beta}\lambda_0^2\right)^2}},
\end{array}
\end{equation}
where $s_{\alpha\beta} = \alpha + \beta$ and $d_{\alpha\beta} = -1 + \alpha\beta$. After the integration with the full response kernel Eq.(\ref{f2}) the result can be written in a more compact form:

\begin{equation}
\label{f11}
\begin{array}{ll}
\displaystyle{ v_{d} = \frac{v_0^2\lambda_0^2 W |\nabla c| \sum_{j=0}^7 {a_j(\alpha, \beta) D_r^{7-j} \lambda_0^j}}{4\sum_{j=0}^{10} {b_j(\alpha,\beta) D_r^{10-j} \lambda_0^j}}},
\end{array}
\end{equation}
where the angle-dependent functions $a_j(\alpha, \beta)$ and $b_j(\alpha, \beta)$ can be found in \hyperref[S1_Appendix]{S1 Appendix}. Note that, from the orders of polynomials and expressions for non-vanishing coefficients (for any $\alpha\in(-1; 1)$ and $\beta\in(-1; 1)$) at the highest powers of $D_r$ and $\lambda_0$ follows that the drift velocity is always inversely proportional to the base-line turning rate $\lambda_0$ and to rotational diffusion coefficient in the third degree, i.e. $v_d\sim1/D_r^3$. Moreover, it is easy to check, that by setting $\beta = 0$ we recover the results of \cite{Taktikos2013} and for $\alpha=\beta$ the expression of Locsei \cite{Locsei2007}. Also as in \cite{Taktikos2013} and \cite{Locsei2007}, for any two alternating turning events the drift velocity is quadratic in bacteria speed $v_0$. This scaling can be qualitatively understood as follows. The length of each run of the cell is proportional to its speed, while the bias in this run is determined by the sensed gradient. During a run, the cell translates the spatial gradient into the temporal concentration gradient where the cell velocity enters as a scaling factor, thus resulting in the overall quadratic dependence of the drift speed on cell velocity. Before applying the general result of Eq.(\ref{f11}) to experimental data we first explore its dependence on parameters and compare to the results of numerical simulations.

\subsection*{Verification of analytically obtained formula for the drift speed by numerical simulations}
\label{S3}

The drift speed is linearly proportional to the gradient of concentration $|\nabla c|$ and to the amplitude of the response $W$. Dependence on rotational diffusion constant and the turning frequency (without the gradient) is more involved. Our main goal is to check the angular 
dependence of the result with numerica experiment.

Due to the specific form of the response function, Eq. (\ref{f2}), we can benefit from the so-called \textquotedblleft{}embedding\textquotedblright{} technique \cite{embeding1,embeding2}. We transform Eq. (\ref{f1}), which is nonlocal in time, into a system of three linear differential equations corresponding to the number of terms in the response function and that are now local in time. Bacteria motility is modeled as a sequence of short steps, during which the bacteria performs a motion by integrating its velocity in time and integrates the chemical concentration of the attractant in space. The velocity vector performs an unbiased rotational diffusion. Every step is completed with a \textquotedblleft{}coin flipping\textquotedblright{}, which decides whether the bacteria should perform a re-orientation or not. With decreasing of the time step $\Delta t$, the discrete process converges and starts to reproduce the  continuous stochastic dynamics described in Section \hyperref[S2]{II}. A more detailed description of  the numerical algorithm is given in \hyperref[S2_Appendix]{S2 Appendix}.	

We model the motility of an ensembel of  $N\approx 0.5 \cdot 10^6$ cells during the time interval $T=2000$ s $\approx 33$ min with the time step $\Delta t=0.01$s were performed 
for the following fixed parameters: $v_0 = 45$ $\mu$ms$^{-1}$ and $\tau _{run} = 0.3$ s\cite{Xie2011}. 
The value of the rotational diffusion constant $D_r$, as well as parameters $\alpha$ and $\beta$, were varied. 
We also performed simulations for different strengths of the linear gradient. 

\begin{figure}[h]
	\begin{center}
		\includegraphics[width=130mm]{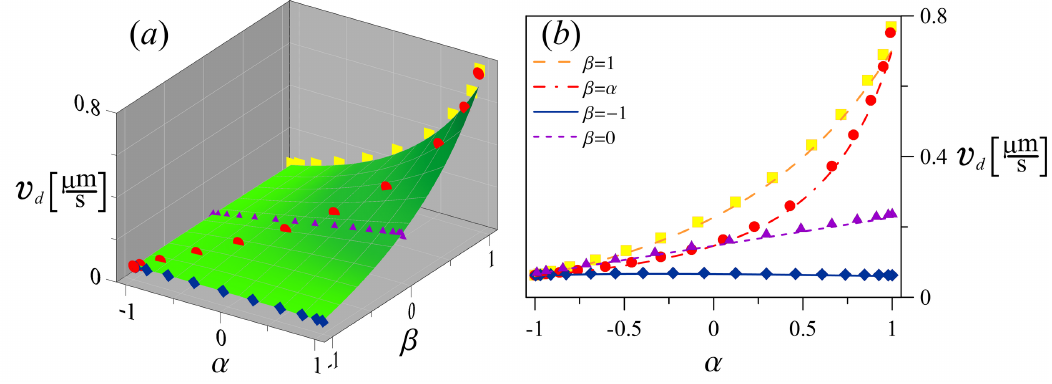}
		\caption{\textbf{Drift speed without rotational diffusion.} \hspace{\textwidth}(a) Analytically obtained function $v_d(\alpha,\beta)$, see Eq.(\ref{f11}) is shown as a green surface and numerically obtained results as symbols for $D_r=0.0$ rad$^2$s$^{-1}$, $|\nabla c|=0.05$ $\mu m^{-4}$ and $W=0.0458$ $\mu$m$^3$. (b) Comparison of analytically (lines) and numerically (symbols) obtained drift speed dependences on the parameter $\alpha$ for four values of $\beta$:  $\beta = 1.0$ (yellow), $\beta = \alpha$ (red), $\beta = -1.0$ (blue), $\beta = 0.0$ (purple).}
		\label{fig2}
	\end{center}
\end{figure}

\begin{figure}[h]
	\begin{center}
		\includegraphics[width=130mm]{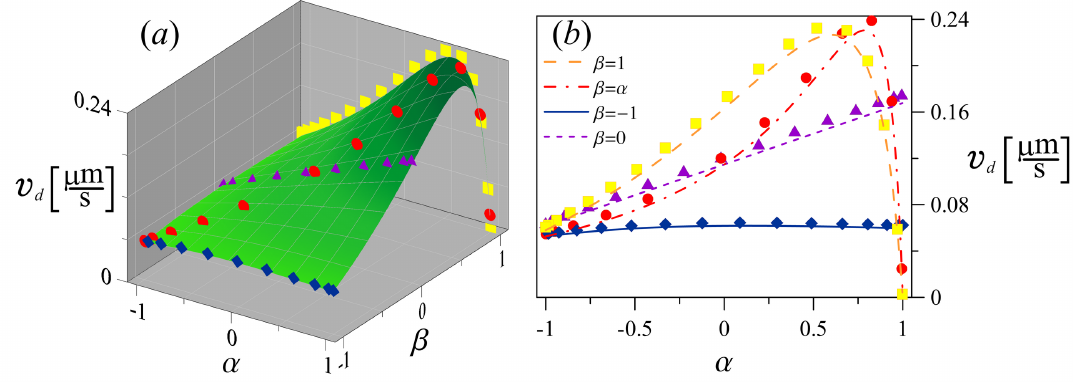}
		\caption{\textbf{Drift speed with rotational diffusion.} \hspace{\textwidth}(a) Analytically obtained function $v_d(\alpha,\beta)$, shown as surface, and numerically obtained result as points for $D_r=0.2$ rad$^2$s$^{-1}$, $|\nabla c| = 0.05$ $\mu m^{-4}$ and $W=0.0458$ $\mu$m$^3$. (b) Comparison of analytically (lines) and numerically (symbols) obtained drift speed dependences on the parameter $\alpha$ for four values of $\beta$:  $\beta = 1.0$ (yellow), $\beta = \alpha$ (red), $\beta = -1.0$ (blue), $\beta = 0.0$ (purple).}
		\label{fig3}
	\end{center}
\end{figure}

For $D_r=0$ rad$^2$s$^{-1}$, analytically obtained drift velocity in the general case of two alternating turning events is presented as a green surface in Fig.~\ref{fig2}. Symbols show the numerically calculated values of the drift speed for three particular types of the swimming patterns: $\beta = 1.0$ (yellow), $\beta = \alpha$ (red), $\beta = -1.0$ (blue), $\beta = 0.0$ (purple). We see that as expected $v_d(\alpha, \beta)$ is symmetric with respect to the plane $\alpha=\beta$. The drift speed is increasing when both $\alpha$ and $\beta$ approach $1$. However when $\alpha=\beta=1$ without rotational diffusion we have directed ballistic motion and $v_d = 0$. Agreement is similarly good for the case $D_r=0.2$ rad$^2$s$^{-1}$, see results on Fig.~\ref{fig3}. We see that the rotational diffusion reduces the drift velocity and also leads to the appearance of a maximum along the line $\alpha=\beta$ with the drift velocity smoothly going to zero as $\alpha$ and $\beta$ approach $1$. 

\begin{figure}[]
	\begin{center}
		\includegraphics[width=100mm]{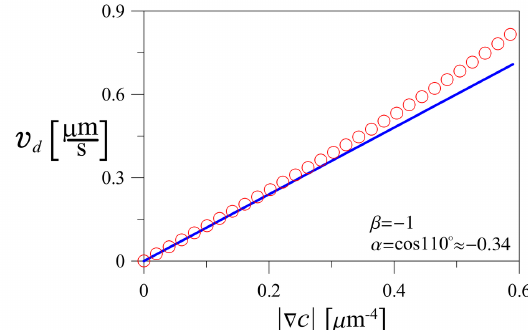}
		\caption{\textbf{Drift speed as a function of the chemoattractant gradient strength.} \hspace{\textwidth}Analytically obtained predictions (curve) agree with numerical results (symbols) up to the gradient values of $\backsimeq 0.5$ $\mu m^{-4}$ ($D_r=0.2$ rad$^2$s$^{-1}$, $\lambda = 3.3$ $\textrm{s}^{-1}$, $\alpha \approx -0.34 , \beta = -1$).}
		\label{fig4}
	\end{center}
\end{figure}

Numerical experiments are consistent with analytical calculations within $10\%$ error up to the gradient value $|\nabla c| \approx 0.6$ $\mu  m^{-4}$, Fig.~\ref{fig4}. This once again highlights the fact, that here we used a linear theory of chemotaxis, which relies on the expansion with respect to a small gradient. Now we can apply our analytical results to the experimental data of \textit{V. alginolyticus} motility.

\section*{Chemotactic drift speed estimation based on experimental data for \emph{V. alginolyticus} swimming pattern}

\begin{figure}[ht!!!]
	\begin{center} 
		\includegraphics[width=130mm]{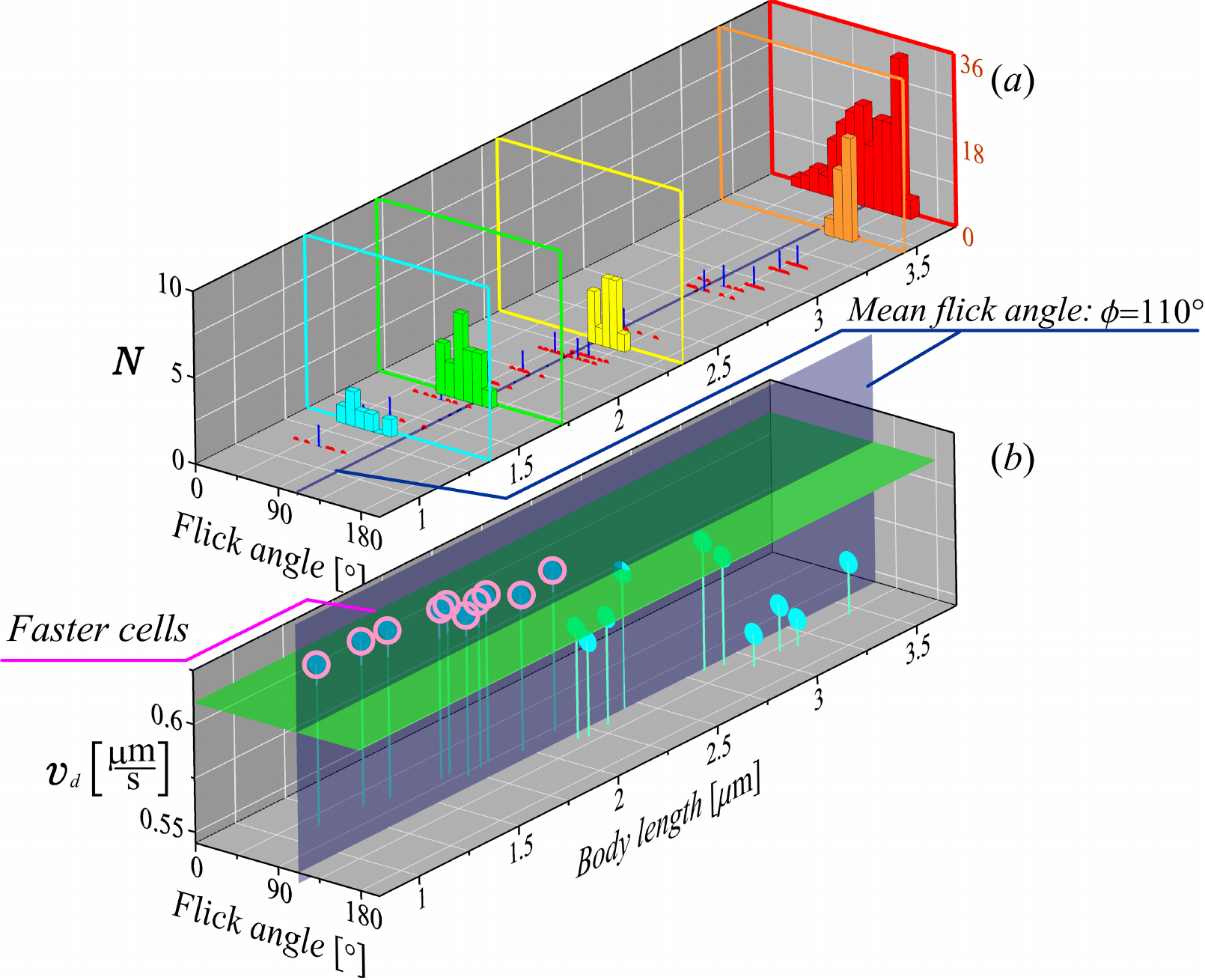}
		\caption{\textbf{Variability of the drift velocity with the cell body size.} \hspace{\textwidth}(a) Histograms showing the distribution of flick angles with various body lengths (data from \cite{Taute2015}). Red histogram represents all measured flick angles -- 170 events for 20 cells with various body lengths. Flick angles for four particular values of cellular body lengths are shown by narrow color histograms ($l=3.45$ $\mu$m, orange; $l=2.31$ $\mu$m, yellow; $l=1.72$  $\mu$m, green; $l=1.35$ $\mu$m, cyan histogram). The data obtained for 20 individuals (each of which is displayed at least 6 flicks in their trajectory) are shown as red points. Small blue ticks show the mean flick angles obtained from individual distributions. Black tick labels of the vertical axis correspond to the histograms of individual cells, whereas the red tick labels correspond to the red histogram  for all cells. (b) Analytically obtained drift speed calculated for the mean flick angle of the considered cells, as a function of the angle and the corresponding body length. For results the following parameters were used: $|\nabla c|=0.5$ $\mu$m$^{-4}$, $\lambda =3.3$ s$^{-1}$, $v_0=45$ $\mu$ms$^{-1}$, $D_r=0.2$ rad$^2$s$^{-1}$, $W=0.0458$ $\mu$m$^3$ and $\beta=\cos{172^\circ}$.}
		\label{fig5}
	\end{center}
\end{figure}

Recently, the trajectories obtained by the high-throughput 3D bacterial tracking method \cite{Taute2015} revealed some interesting details about the \textquotedblleft{}run-reverse-flick\textquotedblright{} swimming pattern of the marine bacterium \textit{V. alginolyticus}. The ability to capture individual trajectories that contain a sufficient number of flicks and the information about the size of the bacteria provided new insights into  inter-individual variability. In Figure \ref{fig5}(a), the experimentally measured distribution of the flick angles is shown as a red histogram. It is rather broad and has one well pronounced maximum. In similar previous measurements \cite{Xie2011}, the flick angles were reported to be randomly distributed with an average turning angle close to $90^{\circ}$. 

Importantly, comparing the measured angles for the cells with different sizes revealed that individuals actually show very narrow flick angle distributions with different 
means [orange, yellow, green and cyan histograms in Fig.~\ref{fig5}(a)]. Moreover, there is a correlation between the cells-body length and the angle between the forward and reverse runs during the flick, i.e. between the cell's size and $\alpha$.  Reorientation during a flick is counteracted by the viscous drag \cite{Son2013} and thus larger cells have a flick angle closer to $180^\circ$ \cite{Taute2015}. 

As was shown in \cite{Taute2015}, the mean value of the reversal angle for the bacteria in the considered population was $172^\circ$. In this case, the reversal has a narrow distribution and almost does not change from cell to cell.

Therefore,  given the more detailed data of \cite{Taute2015}, simplified \textquotedblleft{}run-reverse-flick\textquotedblright{} motility models using the previously measured mean flick angle of $\sim 90^{\circ}$ \cite{Xie2011} and assuming a completely random swimming direction during flicks, should be reconsidered. In this section, we will use the detailed data of \cite{Taute2015} to calculate the drift velocity from our analytical results and show how this velocity varies with the cell size. Specifically, by substituting into Eq.(\ref{f11}) the experimentally measured values of the mean cosines of both turning events, i.e.  $\beta=\cos{172^\circ}\approx -0.99$ for the reversal for all cells and size-dependent $\alpha$, the drift velocities can be calculated for various sizes of cells. This shows that the bacteria having various body lengths and, consequently, various mean flicking angles, demonstrate noticeable variability. The drift velocity obtained for bacteria having small size (up to $2$ $\mu m$) are circled in Fig.~\ref{fig5}(b). This shows that the smaller cells have higher drift speed and, therefore, can reach a chemoattractant source faster than the others, and the difference between slowest and fastest cells is of the order of $10\%$. For results in Fig.~\ref{fig5}(b) we used the following parameters: $|\nabla c|=0.5$ $\mu$m$^{-4}$, $\lambda =3.3$ s$^{-1}$, $v_0=45$ $\mu$ms$^{-1}$, $D_r=0.2$ rad$^2$s$^{-1}$, $W=0.0458$ $\mu$m$^3$ and $\beta=\cos{172^\circ}$. It is important to note, that the value of $W$ we borrowed from \textit{E. coli} bacteria. For \textit{V. alginolyticus} it can be easily back-calculated from the experimentally measured drift speed in a small gradient (as it was done before for \textit{E. coli} \cite{Taktikos2013}). However, to the best of our knowledge, the drift speed of \textit{V. alginolyticus} in a small linear gradient was not measured before. It would be an important next step, also allowing for the experimental validation of our theoretical predictions.

\section*{Discussion}
\label{S4}

Continuously advancing measurements techniques allow us to get a more detailed information on a bacterial behavior at the level of individual cells. We are at the point when the variability between cells can and should be accounted for in our quantitative analysis of motility and chemotaxis. In this work we considered one of the most general swimming patterns containing two alternating turning angles. This analytical framework allows for a straightforward analysis of \textit{V. alginolyticus} cells with their intrinsic variability: cells of different sizes have different flick angles and thus can be naturally accommodated by the model. The conceptual challenge of the provided calculation of the chemotactic drift velocity is in the non-disappearing memory during the turning events. 

\begin{figure}[ht!!!]
	\begin{center}
		\includegraphics[width=130mm]{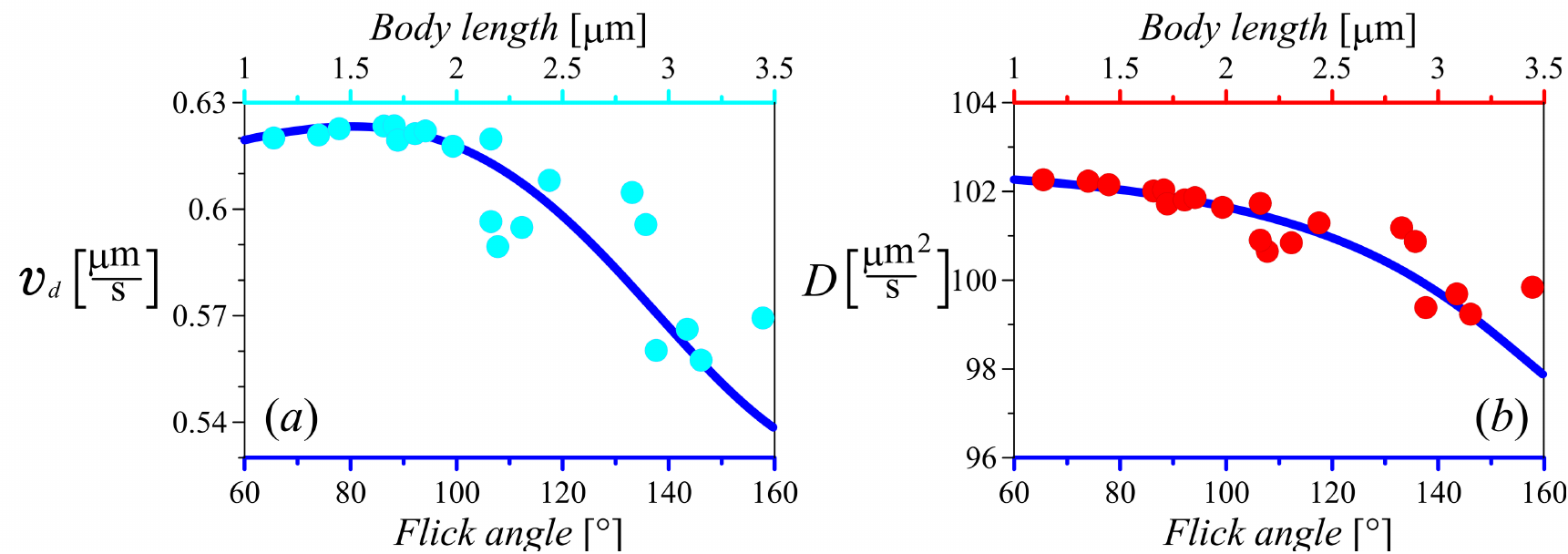}
		\caption{\textbf{Analytically obtained drift speed and diffusion constant as functions of the mean flick angle and the corresponding body length.} \hspace{\textwidth}The curves represent the angle-dependent characteristics obtained from Eqs.(\ref{f11}) and (\ref{f12}). To plot the drift speed and the diffusion constant as a function of cell size, we use the data of \cite{Taute2015} to relate the size to the turning angle, and use that angle in analytical results Eqs.(\ref{f11}) and (\ref{f12}) (these values are shown by symbols). The fact that data based on cell sizes line up with the theoretical curves as functions of angles indicates an approximately linear relation between the cell size and the cosine of the turning angle, which is in agreement with data presented in \cite{Taute2015}. The parameters are $\lambda =3.3$ s$^{-1}$, $v_0=45$ $\mu$ms$^{-1}$, $D_r=0.2$ rad$^2$s$^{-1}$ and $\beta=\cos{172^\circ}$.}
		\label{fig6}
	\end{center}
\end{figure}

Although we observe a noticeable difference in the drift speeds of bacteria of different sizes, for the considered small gradients this difference is of the order of 10 $\%$, see Fig.~\ref{fig6}(a). We should note that although there is a maximum of the drift speed at a certain flick angle, which depends on the parameters of the swimming pattern in a nontrivial way, this maximum is not very pronounced. Thus bacteria with a rather broad range of flick angles have comparable drift velocities. Interestingly, a much stronger dependence on the second turning angle was recently reported for {\em S. putrefaciens} bacterium \cite{Bubendorfer2014}. This effect has a natural explanation. In {\em S. putrefaciens} the duration of the run after reversal and prior to flick is much shorter than that of the run after the flick. Thus the motility pattern is effectively similar to the {\em E. coli} run and tumble with a single turning angle. In that case the effect of the angle on the drift velocity is very pronounced (see Fig.~\ref{fig3}(b) red curve). Importantly, further customization of the model is possible. We can consider different (but limited to exponential) distributions for two run times, different forward and backward speeds, and even different memory kernels. Thus the example of  {\em S. putrefaciens}  can be also put on the analytical footing developed in this paper.

Another important parameter affecting the drift speed is the rotational diffusion (cf. Figs. \ref{fig2} and \ref{fig3}). Potentially, rotational diffusion is another parameter that can depend on the cell size and it would be important to check this effect both theoretically and experimentally in the future.

For bacterial chemotaxis it is not only important how fast cells can move towards the higher concentration of signaling chemicals, but also how well can they localize themselves near the source of the gradient. Thus not only the drift velocity but also the effective diffusion constant of the bacterial motility pattern play an important role \cite{Xie2014}. One can quantify the localization ability by considering for example the ratio $v_d/D$, which describes the competition between the drift and diffusive spreading and has the meaning of the inverse characteristic length. The diffusion constant of the motility pattern with two arbitrary turning angles was calculated in \cite{Taktikos2013}: 

\begin{equation}
\label{f12}
D = \frac{v_0^2}{6}\frac{(2+\alpha+\beta)\lambda+4D_r}{(1-\alpha\beta)\lambda^2+4D_r (\lambda+D_r)}.
\end{equation}

As for the drift velocity, substituting into Eq.(\ref{f12}) the mean cosines of alternating size-dependent flicks $\alpha$ and size-independent reversal angles $\beta$, one can estimate the impact of bacterial body length in cell's localization ability near the source of attracting chemicals. It is easy to see that in general there is a strong dependence of the diffusion constant on the turning angles. However, if one of the angles is fixed in the reversal mode, as in the case of {\em V.alginolyticus}, the variation of the flick angle does not lead to large changes in the effective diffusion constant, see Fig.~\ref{fig6}(b) (for a smaller rotational diffusion constant the effect of the flick angle would be more pronounced). Note that, as for the drift velocity, the diffusion coefficient is also quadratic in bacteria speed $v_0$. This scaling is due to a simple random walk estimate of the diffusion constant as the mean squared step distance $(v_0\tau_{run})^2$ divided by the mean run time $\tau_{run}$. 

With this information at hand we have a full theoretical tool set to quantify bacterial chemotaxis in the small gradient approximation. The theoretical results presented here for a 4-step pattern, allow us to predict the drift as a function of the cell size. Importantly, these predictions could be verified experimentally where the drift speeds of cells along the linear gradient can be correlated with their size. One of the possible applications of the size-dependent drift velocity is in cell sorting, where after a certain time of motion along the linear gradient the cells of different sizes would be found at distinct positions corresponding to their drift speed.

We think that the analytical relations, such as chemotactic drift speed considered here, provide a rigorous link between motility pattern and chemotactic response and thus can be used in experiments to infer the yet unknown parameters of bacterial sensitizing based on tracking or chemotactic drift experiments.

\section*{Acknowledgments}
This work was supported by RSF project 16-12-10496.

\section*{Supporting information}

\appendix
\renewcommand{\theequation}{S\arabic{equation}}
\setcounter{equation}{0}
\paragraph*{S1 Appendix.}
\label{S1_Appendix}
{\bf Drift speed calculation}

The motility patterns of bacteria can be considered as a process composed of two main alternating phases of motion: \textquotedblleft{}runs\textquotedblright{} and \textquotedblleft{}turnings\textquotedblright{}. Although the angular changes during the turning events are stochastic, analysis of the experimental data shows that cells are able to exhibit directional persistence that can be different for various species. \textit{E. coli} cells have a motility pattern with a preferential turning angle $\Delta\varphi$ that can be characterized by the parameter $\alpha = \langle\cos{\Delta\varphi}\rangle$. \textit{V. alginolyticus} bacteria perform motility with two 
alternating re-orientation events that can be specified by two parameters $\alpha = \langle\cos{\Delta\varphi_1}\rangle$ and $\beta = \langle\cos{\Delta\varphi_2}\rangle$.

The calculation of chemotactic drift speed $v_d$ for motility with a single directional parameter was performed by Locsei \cite{Locsei2007}. Here, we extend his approach to more complicated case of bacterial chemotaxis with two alternating turning events. Note that, the algorithm presented below can be easily extended for random walks with arbitrary number of alternating turning events. Moreover, taking into account the change of parameters (mean run durations and rotational diffusion coefficients) between appointed turnings, more accurate prediction for the chemotactic drift speed can be obtained. Here, for simplicity, we restrict our analysis by assumption that all parameters for \textquotedblleft{}run\textquotedblright{}-phases of the considered pattern are the same.

Without loss of generality we consider a random walk whose first turning event for $t = t_{\beta} > 0$ is specified by the parameter $\beta$, i.e. with a turning of type $\alpha$ at $t=0$.

Let $z_{\beta}$ be the $z$ location of a cell at the end of a run, relative to its position at the beginning of the considered random walk. At $t=t_{\beta}$, the direction of motion is changed with parameter $\beta$ and the cell performs a run of duration $t_{\alpha}$. To obtain the drift velocity $v_d$ in a first-order expansion with respect to $|\nabla c|$, we determine the average displacement of a run $\langle \overline{z}_{\beta} \rangle$ and a subsequent run $\langle \overline{z}_{\alpha} \rangle$. In our calculations, as in \cite{Locsei2007}, we treat the duration of turning events as negligible. As the expressions for $\langle \overline{z}_{\beta} \rangle$ and $\langle \overline{z}_{\alpha} \rangle$ are first order in $|\nabla c|$, the mean duration of \textquotedblleft{}two runs\textquotedblright{} process is given by $2\tau_{run}$, and the chemotaxis drift speed becomes

\begin{equation}
	\label{eq1}
	v_d = \frac{\langle \overline{z}_{\beta} \rangle + \langle \overline{z}_{\alpha} \rangle}{2 \tau_{run}}.
\end{equation}

For $\langle \overline{z}_{\alpha} \rangle$ and $\langle \overline{z}_{\beta} \rangle$ we calculate expectation over all possible paths taking into account that the position of the bacterium at a time $t$ on a particular path is random. In the first step, we calculate the mean displacement of the run $\langle \overline{z}_{\beta} \rangle$:

\begin{equation}
	\label{eq2}
	\langle \overline{z}_{\beta} \rangle = \langle \int_{0}^{\infty} z(t) p(t) dt \rangle.
\end{equation}
where $z(t)$ is position of a cell at time $t$ in particular path, and $p(t)$ is the probability density, that a run starts at $t=0$ and stops and $t=t_{\beta}$. Since the paths are independent of $t_{\beta}$, we may take the path expectation inside the integral over run times

\begin{equation}
	\label{eq3}
	\langle \overline{z}_{\beta} \rangle = \int_{0}^{\infty} \langle  z(t) p(t) \rangle dt,
\end{equation}
where $p(t)$ is given by \cite{Locsei2007}

\begin{equation}
	\label{eq4}
	p(t) = \lambda(t) e^{-\int_{0}^{t}\lambda(t')dt'}.
\end{equation}

In accordance with the idea of de Gennes we assume that the turning rate $\lambda\left( t \right)$ in (\ref{eq4}) in the presence of chemoattractant becomes 

\begin{equation}
	\label{eq5}
	\lambda(t) = \lambda_0 [1-\Delta(t)],
\end{equation}
where $\lambda_0 = 1 / \tau_{run}$ is the mean turning rate of bacteria, and the fractional change in $\lambda \left( t \right)$ caused by the presence of chemicals in the environment, is:

\begin{equation}
	\label{eq6}
	\Delta(t) = \int_{-\infty}^{t} c(t') R(t-t') dt'.
\end{equation}
In (\ref{eq6}) $c(t')$ is chemoattractant concentration experienced by the cell at time $t'$ and $R(t)$ is the cell's response function.
In the case of small chemical gradient $|\nabla c|$ in $z$ direction, the chemoattractant concentration $c(t)$ at the cell's position $z(t)$ can be written in the form

\begin{equation}
	\label{eq7}
	c(t) = c_0 + |\nabla c| z(t).
\end{equation}
Substituting (\ref{eq7}) into (\ref{eq6}) yields

\begin{equation}
	\label{eq8}
	\Delta(t) = \int_{-\infty}^{t} [c_0 + |\nabla c| z(t')] R(t-t') dt' = c_0 \int_{-\infty}^{t} R(t-t')dt' + |\nabla c|\int_{-\infty}^{t} z(t')R(t-t') dt'.
\end{equation}
Since the response function $R$ has a zero integral, an additive constant in (\ref{eq8}) has no effect on $\Delta (t)$. Therefore, the equality (\ref{eq8}) can be rewritten in the form

\begin{equation}
	\label{eq9}
	\Delta(t) = |\nabla c|\int_{-\infty}^{t} z(t')R(t-t') dt'.
\end{equation}

To simplify forthcoming calculations, we first consider the special case of a delta-response in time

\begin{equation}
	\label{eq10}
	R(t) = A \delta(t-T)
\end{equation}
with delay time $T$ and strength $A$, as it was done in \cite{deGennes2004}. In this case, the fractional change in turning rate becomes

\begin{equation}
	\label{eq11}
	\Delta(t) = A|\nabla c|\int_{-\infty}^{t} z(t')\delta(t-T-t') dt' = A |\nabla c| z(t-T).
\end{equation}
Thus, taking into account that for the run lasting from $t=0$ until $t=t_{\beta}$, the average displacement (\ref{eq3}) after integration by parts can be written as 

\begin{equation}
	\label{eq12}
	\langle \overline{z}_{\beta} \rangle = \int_{0}^{\infty} \langle v_z (t) e^{-\int_{0}^{t}\lambda\left( t ' \right) dt'} \rangle dt,
\end{equation}
where $v_z \left(t\right) = dz\left(t\right) / dt$ and $\lambda\left(t'\right) = \lambda_0 \left[1-A|\nabla c|z\left(t'-T\right)\right]$, one can obtain 

\begin{equation}
	\label{eq13}
	\langle \overline{z}_{\beta} \rangle = \int_{0}^{\infty} \langle v_z (t) e^{-\lambda_0 t} e^{\lambda_0 A|\nabla c|\int_{0}^{t}z\left(t'-T\right)dt'} \rangle dt.
\end{equation}

For a small chemical gradient ($\Delta(t) \ll 1$) expanding the exponential and keeping only the first-order terms in $|\nabla c|$, i.e. $e^{\lambda_0 A |\nabla c| \int_{0}^{t} z(t'-T) dt'} \approx 1 + \lambda_0 A |\nabla c| \int_{0}^{t} z(t'-T) dt'$, one obtains

\begin{equation}
	\label{eq14}
	\langle \overline{z}_{\beta} \rangle = 
	\int_{0}^{\infty} \langle v_z (t) \rangle e^{-\lambda_0 t} dt + \lambda_0 A |\nabla c| \int_{0}^{\infty} e^{-\lambda_0 t} \left[  \int_{0}^{t} \langle z(t'-T) v_z(t) \rangle dt' \right] dt.
\end{equation}
Substitution of $\langle v_z(t) \rangle = e^{-2D_r t} \langle v(0^+) \rangle$ \cite{Locsei2007} into the first integral of (\ref{eq14}) and using $z(t) = \int_{0}^{t} v_z(s) ds$ for the second integral of (\ref{eq14}) yields

\begin{equation}
	\label{eq15}
	\langle \overline{z}_{\beta} \rangle = \frac{\langle v(0^+) \rangle}{\lambda_0 + 2D_r} + \lambda_0 A |\nabla c| \int_{0}^{\infty} e^{-\lambda_0 t} \left[ \int_{0}^{t} \left( \int_{0}^{t'-T} \langle v_z(s) v_z(t) \rangle ds \right) dt' \right] dt.
\end{equation}

Also within the first order approximation, the velocity is governed by an isotropic distribution, which implies

\begin{equation}
	\label{eq16}
	\langle v_z(s) v_z(t_{\beta}) \rangle = \frac{v_0^2}{3} \langle \mathbf{e}(s)\cdot\mathbf{e}(t_{\beta}) \rangle
\end{equation}
where $v_0$ is a constant speed during a run, and the directional correlation function reads \cite{Taktikos2013}

\begin{equation}
	\label{eq17}
	\langle \mathbf{e}(s) \cdot \mathbf{e}(t_{\beta})\rangle = \left\{
	\begin{array}{ll}
		e^{-2 D_r (t_{\beta}-s)} & \textrm{if } 0 \le s < t_{\beta} \textrm{,}\\
		\alpha e^{-2D_r t_{\beta}} e^{-(\lambda_0+2D_r) |s|} \cdot \\
		\left[ \sqrt{\frac{\beta}{\alpha}} \sinh{\left(\sqrt{\alpha}\sqrt{\beta}\lambda_0 |s|\right)} + \cosh{\left(\sqrt{\alpha}\sqrt{\beta}\lambda_0 |s|\right)} \right] & \textrm{if } s<0.
	\end{array} \right.
\end{equation}
To obtain the second line of (\ref{eq17}), we made a decomposition

\begin{equation}
	\label{eq18}
	\langle \mathbf{e}(s) \cdot \mathbf{e}(t_{\beta})\rangle = \langle \mathbf{e}(s) \cdot \mathbf{e}(0^-)\rangle \langle \mathbf{e}(0^-) \cdot \mathbf{e}(0^+)\rangle \langle \mathbf{e}(0^+) \cdot\mathbf{e}(t_{\beta})\rangle,
\end{equation}
where the direction correlation function for directions immediately before and after the turning at $t=0$ equals to $\alpha$ due to our choice of the considered motility pattern:

\begin{equation}
	\label{eq19}
	\langle \mathbf{e}(0^-) \cdot\mathbf{e}(0^+)\rangle = \alpha,
\end{equation}
whereas the last multiplier in (\ref{eq18}) can be obtained from the Fokker-Planck equation (see \hyperref[S2_Appendix]{S2 Appendix} and \cite{Locsei2007} for details).

\begin{equation}
	\label{eq20}
	\langle \mathbf{e}(0^+) \cdot\mathbf{e}(t_\beta)\rangle = e^{-2D_r t_\beta},
\end{equation}
where $D_r$  is the rotational diffusion coefficient  describing the influence of rotational Brownian motion during the \textquotedblleft{}run\textquotedblright{}-phase of bacterial swimming. To determine the first multiplier in the product (\ref{eq18}) we should take into account that for $t<0$ the cell's motility obeys to the pattern with two alternating turning events with additional reorientations due to rotational Brownian motion during the runs. For this case, the direction correlation function for the process, whose first run interrupts by a turning of type $\beta$ becomes 

\begin{equation}
	\label{eq21}
	\langle \mathbf{e}(s) \cdot\mathbf{e}(0^-)\rangle = e^{-(\lambda_0 +2D_r)|s|} \left[ \sqrt{\frac{\beta}{\alpha}} \sinh{\left(\sqrt{\alpha}\sqrt{\beta}\lambda_0 |s|\right)} + \cosh{\left(\sqrt{\alpha}\sqrt{\beta}\lambda_0 |s|\right)} \right],
\end{equation}
see \cite{Taktikos2013} for details. After inserting (\ref{eq16}) and (\ref{eq17}) into the second integral of (\ref{eq15}) for the mean displacement of the first run we obtain

\begin{equation}
	\label{eq22}
	\langle \overline{z}_{\beta} \rangle = \frac{\langle v(0^+) \rangle}{\lambda_0 + 2D_r} + \frac{v_0^2}{3} \lambda_0 A |\nabla c| \left[k_{\beta}\cosh{\left(\sqrt{\alpha}\sqrt{\beta}\lambda_0 T\right)} + m_{\beta}\sinh{\left(\sqrt{\alpha}\sqrt{\beta}\lambda_0 T\right)} + n_{\beta}\right],
\end{equation}
where

\begin{equation}
	\label{eq23}
	\begin{array}{ll}
		\displaystyle{k_{\beta} =
			\frac{\lambda_0 e^{-(2D_r+\lambda_0)T}(2D_r+\lambda_0+\alpha\lambda_0)}{(2D_r+\lambda_0) (4D_r^2+4D_r\lambda_0+(1-\alpha\beta)\lambda_0^2)}}, \\
		\displaystyle{m_{\beta} = \frac{\lambda_0 e^{-(2D_r+\lambda_0)T}\sqrt{\alpha}(2D_r+\lambda_0+\beta\lambda_0)}{\sqrt{\beta}(2D_r+\lambda_0) (4D_r^2+4D_r\lambda_0+(1-\alpha\beta)\lambda_0^2)}}, \\
		\displaystyle{n_{\beta} =
			\frac{-\lambda_0^2\alpha(2D_r+\lambda_0+\beta\lambda_0)}{(2D_r+\lambda_0)^2 (4D_r^2+4D_r\lambda_0+(1-\alpha\beta)\lambda_0^2)}}.
	\end{array}
\end{equation}

Proceeding in the same way as for $z_{\beta}$, one can obtain the mean displacement of the second run, which is interrupted by turning of type $\alpha$:

\begin{equation}
	\label{eq24}
	\langle \overline{z}_{\alpha} \rangle = \frac{\langle v(t_\beta^+) \rangle}{\lambda_0 + 2D_r} + \frac{v_0^2}{3} \lambda_0 A |\nabla c| \left[k_{\alpha}\cosh{\left(\sqrt{\alpha}\sqrt{\beta}\lambda_0 T\right)} + m_{\alpha}\sinh{\left(\sqrt{\alpha}\sqrt{\beta}\lambda_0 T\right)} + n_{\alpha}\right],
\end{equation}
where

\begin{equation}
	\label{eq25}
	\begin{array}{ll}
		\displaystyle{k_{\alpha} =
			\frac{\lambda_0 e^{-(2D_r+\lambda_0)T}(2D_r+\lambda_0+\beta\lambda_0)}{(2D_r+\lambda_0) (4D_r^2+4D_r\lambda_0+(1-\alpha\beta)\lambda_0^2)}}, \\
		\displaystyle{m_{\alpha} = \frac{\lambda_0 e^{-(2D_r+\lambda_0)T}\sqrt{\beta}(2D_r+\lambda_0+\alpha\lambda_0)}{\sqrt{\alpha}(2D_r+\lambda_0) (4D_r^2+4D_r\lambda_0+(1-\alpha\beta)\lambda_0^2)}}, \\
		\displaystyle{n_{\alpha} =
			\frac{-\lambda_0^2\beta(2D_r+\lambda_0+\alpha\lambda_0)}{(2D_r+\lambda_0)^2 (4D_r^2+4D_r\lambda_0+(1-\alpha\beta)\lambda_0^2)}}
	\end{array}
\end{equation}
and $v(t_\beta^+)$ is the $z$ component of the cell's velocity at the beginning of the second run. As before, to obtain the directional correlation function for $t<t_\alpha$, we should expand $\langle \mathbf{e}(s) \cdot \mathbf{e}(t_{\alpha})\rangle$ into a product of direction correlations between different times. However, since we consider the motility with regularly alternating reorientations of bacteria, we can claim its similarity to the previous one (\ref{eq17}) with obvious replacement for the following parameters $t_{\beta} \rightarrow t_{\alpha}$ and $\alpha \leftrightarrow \beta$.

Having found the integrals in expressions for $\langle\overline{z}_\beta \rangle$ and $\langle\overline{z}_\alpha \rangle$, we turn our attention to $\langle v(0^+) \rangle$ and $\langle v(t_\beta^+) \rangle$. In our case, for random walk with a turning of type $\alpha$ at $t=0$ and type $\beta$ at $t=t_\beta$, using the definitions for the parameters $\alpha$ and $\beta$ we can write

\begin{equation}
	\label{eq26}
	\begin{array}{ll}
		\langle \textbf{e}(0^-) \cdot \textbf{e}(0^+) \rangle = \alpha, \\
		\langle \textbf{e}(t_\beta^-) \cdot \textbf{e}(t_\beta^+) \rangle = \beta.
	\end{array}
\end{equation}

From these definitions follows that the expected velocities immediately after the turning event can be expressed in terms of the expected velocities immediately before it (see \cite{Locsei2007} for details):

\begin{equation}
	\label{eq27}
	\begin{array}{ll}
		\langle v(0^+) \rangle = \alpha \langle v(0^-) \rangle, \\
		\langle v(t_\beta^+) \rangle = \beta \langle v(t_\beta^-) \rangle.
	\end{array}
\end{equation}
On the other hand, to get the expected velocity at the end of run commencing at $t=0$ and interrupting at $t=t_{\beta}$ we may take the integral

\begin{equation}
	\label{eq28}
	\langle v(t_\beta^-) \rangle = \int_{0}^{\infty} \langle  v(t) p(t) \rangle dt.
\end{equation}
Substituting (\ref{eq5}) into (\ref{eq4}) and using (\ref{eq11}) yields

\begin{equation}
	\label{eq29}
	p(t) = \lambda_0 \left[ 1-A|\nabla c| z(t-T) \right] e^{-\lambda_0 \int_{0}^{t} \left[ 1-A|\nabla c| z(t'-T) \right] dt'} .
\end{equation}
As before, expanding the exponential and keeping only the linear term, we can obtain

\begin{equation}
	\label{eq30}
	p(t) = \lambda_0 e^{-\lambda_0 t} \left[ 1-A|\nabla c|z(t-T)+\lambda_0A|\nabla c|\int_{0}^{t}z(t'-T) dt' \right].
\end{equation}
According to (\ref{eq27}), we obtain the expression for $\langle v(t_{\beta}^+) \rangle$ 

\begin{equation}
	\label{eq31}
	\langle v(t_\beta^+) \rangle = \beta \left[\frac{\langle v(0^+)\rangle\lambda_0}{\lambda_0+2D_r}-\lambda_0A|\nabla c|(I_1^{(1)} - I_2^{(1)})\right].
\end{equation}
where

\begin{equation}
	\label{eq32}
	I_1^{(1)} = \int_{0}^{\infty} e^{-\lambda_0 t} \langle v(t)z(t-T) \rangle dt,
\end{equation}
and

\begin{equation}
	\label{eq33}
	I_2^{(1)} = \lambda_0\int_{0}^{\infty} \left( \int_{t'}^{\infty} e^{-\lambda_0 t} \langle v(t)z(t'-T) \rangle dt \right) dt'.
\end{equation}
Similar calculations for the second run give us the expression for the expected velocity immediately before the turning of type $\alpha$ at $t=t_\alpha+t_\beta$

\begin{equation}
	\label{eq34}
	\langle v(t_\alpha^-) \rangle = \frac{\langle v(t_\beta^+)\rangle\lambda_0}{\lambda_0+2D_r}-\lambda_0A|\nabla c|(I_1^{(2)} - I_2^{(2)}).
\end{equation}
But, since the expected velocity before this $\alpha$-turning does not change from that was for $\alpha$-turning at $t=0$, i.e. $\langle v(t_\alpha^-)\rangle=\langle v(0^-)\rangle$, and since $\langle v(0^+)\rangle=\alpha\langle v(0^-)\rangle$, the equation (\ref{eq34}) can be rewritten as 

\begin{equation}
	\label{eq35}
	\langle v(0^+) \rangle = \alpha \left[\frac{\langle v(t_\beta^+)\rangle\lambda_0}{\lambda_0+2D_r}-\lambda_0A|\nabla c|(I_1^{(2)} - I_2^{(2)})\right].
\end{equation}
The equalities (\ref{eq31}) and (\ref{eq35}) give us a system of linear equations with variables $\langle v(0^+) \rangle$ and $\langle v(t_\beta^+)\rangle$ that provide expressions for the cell's velocities at the beginning of run after type $\alpha$ and type $\beta$ turning events, respectively

\begin{equation}
	\label{eq36}
	\begin{array}{ll}
		\displaystyle{\langle v(0^+) \rangle = \frac{\alpha A|\nabla c| \lambda_0 \left( \lambda_0 + 2D_r \right)^2}{\left( \lambda_0 + 2D_r \right)^2 - \alpha\beta\lambda_0^2} \left[ \frac{\beta\lambda_0}{\lambda_0 + 2D_r} \left(I_2^{(1)} - I_1^{(1)}\right) - I_1^{(2)} + I_2^{(2)}\right]},  \\
		\displaystyle{\langle v(t_\beta^+) \rangle = \frac{\beta A|\nabla c|\lambda_0 \left( \lambda_0 + 2D_r \right)^2}{\left( \lambda_0 + 2D_r \right)^2 - \alpha\beta\lambda_0^2} \left[ \frac{\alpha\lambda_0}{\lambda_0 + 2D_r} \left(I_2^{(2)} - I_1^{(2)}\right) - I_1^{(1)} + I_2^{(1)}\right]}.
	\end{array}
\end{equation}

Finally, we combine all preceding results according to (\ref{eq1}), and thus arrive at the chemotactic drift speed for the delta-response

\begin{equation}
	\label{eq37}
	v_\delta = \frac{v_0^2}{3} \lambda_0 A|\nabla c| \left[ k_\delta \cosh{\left(\sqrt{\alpha}\sqrt{\beta}\lambda_0 T\right)} + m_\delta \sinh{\left(\sqrt{\alpha}\sqrt{\beta}\lambda_0 T\right)} + n_\delta \right]
\end{equation}
where denoting $s_{\alpha\beta} = \alpha + \beta$ and $d_{\alpha\beta} = -1 + \alpha\beta$ the coefficients $k_\delta$, $m_\delta$, $n_\delta$ are: 

\begin{equation}
	\label{eq38}
	\begin{array}{ll}
		\displaystyle{k_\delta = \frac{\lambda_0 e^{-\left(2D_r+\lambda_0\right)T} \left( 4D_r^2\left(2-s_{\alpha\beta}\right) - 8D_r d_{\alpha\beta}\lambda_0 - \left(2+s_{\alpha\beta}\right)d_{\alpha\beta}\lambda_0^2 \right)}{2\left(4D_r^2+4D_r\lambda_0-d_{\alpha\beta}\lambda_0^2\right)^2}}, \\ 
		
		\displaystyle{m_\delta = \frac{\lambda_0 e^{-\left(2D_r+\lambda_0\right)T} \left( 4D_r^2\left(s_{\alpha\beta}-2\alpha\beta\right) - 4D_rd_{\alpha\beta}\lambda_0 s_{\alpha\beta} - \lambda_0^2d_{\alpha\beta}\left(s_{\alpha\beta}+2\alpha\beta\right) \right)}{2\left(4D_r^2+4D_r\lambda_0-d_{\alpha\beta}\lambda_0^2\right)^2 \sqrt{\alpha}\sqrt{\beta}}}, \\
		
		\displaystyle{n_\delta = \frac{\lambda_0^2 \left( 4D_r^2\left(-s_{\alpha\beta}+2\alpha\beta\right) + 4D_r\lambda_0d_{\alpha\beta}s_{\alpha\beta} + \lambda_0^2d_{\alpha\beta}\left(s_{\alpha\beta}+2\alpha\beta\right) \right)} {2\left(2D_r+\lambda_0\right)\left(4D_r^2+4D_r\lambda_0-d_{\alpha\beta}\lambda_0^2\right)^2}},
	\end{array}
\end{equation}

Having found $v_\delta$, we obtain the chemotaxis drift speed for the response function

\begin{equation}
	\label{eq39}
	R(T) = W \lambda_0 e^{-\lambda_0 T} \left[ 1-\frac{\lambda_0 T}{2} - \left( \frac{\lambda_0 T}{2} \right)^2 \right],
\end{equation}
where $W$ is a single normalization constant with the dimension of volume, according to

\begin{equation}
	\label{eq40}
	v_{d} = \int_{0}^{\infty} R(T) \frac{v_\delta (t)}{A} dT,
\end{equation}
that yields

\begin{equation}
	\label{eq41}
	\begin{array}{ll}
		\displaystyle{ v_{d} = \frac{v_0^2\lambda_0^2 W |\nabla c| \sum_{j=0}^7 {a_j(\alpha, \beta) D_r^{7-j} \lambda_0^j}}{4\sum_{j=0}^{10} {b_j(\alpha, \beta) D_r^{10-j} \lambda_0^j}}},
	\end{array}
\end{equation}
where 
\begin{flalign*}
	&\displaystyle{a_0(\alpha, \beta) = - 256\left(-2+s_{\alpha\beta}\right)}, \\
	&\displaystyle{a_1(\alpha, \beta) = - 64\left(-46+12\alpha\beta+17s_{\alpha\beta}\right)}, \\
	&\displaystyle{a_2(\alpha, \beta) = - 32\left(-222+116\alpha\beta+(51+2\alpha\beta)s_{\alpha\beta}\right)}, \\
	&\displaystyle{a_3(\alpha, \beta) = 16\left( 582-448\alpha\beta+20\alpha^2\beta^2-(48+29\alpha\beta)s_{\alpha\beta} \right)}, \\
	&\displaystyle{a_4(\alpha, \beta) = 8\left( 894-896\alpha\beta+108\alpha^2\beta^2+(63-126\alpha\beta+10\alpha^2\beta^2)s_{\alpha\beta} \right)}, \\
	&\displaystyle{a_5(\alpha, \beta) = - 4\left( -804+994\alpha\beta-222\alpha^2\beta^2+4\alpha^3\beta^3+(-183+250\alpha\beta-53\alpha^2\beta^2)s_{\alpha\beta} \right)}, \\
	&\displaystyle{a_6(\alpha, \beta) = - 2\left( -392+590\alpha\beta-194\alpha^2\beta^2-4\alpha^3\beta^3+(-152+243\alpha\beta-97\alpha^2\beta^2+6\alpha^3\beta^3)s_{\alpha\beta} \right)}, \\
	&\displaystyle{a_7(\alpha, \beta) = - d_{\alpha\beta}^2\left( -80-12\alpha\beta+4\alpha^2\beta^2+(-44+7\alpha\beta)s_{\alpha\beta} \right)},&
\end{flalign*}
and
\begin{flalign*}
	&\displaystyle{b_0(\alpha, \beta) = 1024}, \\
	&\displaystyle{b_1(\alpha, \beta) = 8192}, \\
	&\displaystyle{b_2(\alpha, \beta) = 29184 - 1280\alpha\beta}, \\
	&\displaystyle{b_3(\alpha, \beta) = 60928 - 8192\alpha\beta}, \\
	&\displaystyle{b_4(\alpha, \beta) = 82496 - 22784\alpha\beta + 640\alpha^2\beta^2}, \\
	&\displaystyle{b_5(\alpha, \beta) = 75648 - 35968\alpha\beta + 3072\alpha^2\beta^2}, \\
	&\displaystyle{b_6(\alpha, \beta) = 47552 - 35248\alpha\beta + 6144\alpha^2\beta^2 - 160\alpha^3\beta^3}, \\
	&\displaystyle{b_7(\alpha, \beta) = 20224 - 21952\alpha\beta + 6560\alpha^2\beta^2 - 512\alpha^3\beta^3}, \\
	&\displaystyle{b_8(\alpha, \beta) = 5568 - 8480\alpha\beta + 3948\alpha^2\beta^2 - 624\alpha^3\beta^3 + 20\alpha^4\beta^4}, \\
	&\displaystyle{b_9(\alpha, \beta) = 896 - 1856\alpha\beta + 1272\alpha^2\beta^2 - 344\alpha^3\beta^3 + 32\alpha^4\beta^4}, \\
	&\displaystyle{b_{10}(\alpha, \beta) = 64 - 176\alpha\beta + 172\alpha^2\beta^2 - 73\alpha^3\beta^3 + 14\alpha^4\beta^4 - \alpha^5 \beta^5}.&
\end{flalign*}

If we assume, that all turning angles are equal ($\alpha = \beta$), we obtain the following expression

\begin{equation}
	\label{eq42}
	v_{d} = W |\nabla c| v_0^2 \frac{\lambda_0^2 (1-\beta)(4D_r+\lambda_0(5-2\beta))}{6(2D_r+\lambda_0-\beta\lambda_0) (2D_r+(2-\beta)\lambda_0)^3},
\end{equation}
which agrees with formula (27) in \cite{Taktikos2013}. If we assume, that the cell's motion pattern is \textquotedblleft{}run-tumble-flick\textquotedblright{} ($\alpha = 0$), we get

\begin{equation}
	\label{eq43}
	\begin{array}{ll}
		\displaystyle v_{d} = \frac{W |\nabla c| v_0^2 \lambda_0^2}{192(D_r+\lambda_0)^4(2D_r+\lambda_0)^2} \cdot \\
		(16D_r^3(2-\beta) + 4D_r^2\lambda_0(22-5\beta) + 2D_r\lambda_0^2(38+5\beta) + \lambda_0^3(20+11\beta)),
	\end{array}
\end{equation}
which also agrees with formula (28) in \cite{Taktikos2013}.

\appendix
\paragraph*{S2 Appendix.}
\label{S2_Appendix}
{\bf Simulation algorithm}

To verify the obtained analytical results, we performed  numerical sampling over an ensemble of chemotaxis trajectories. 
In this section we describe  the used simulation algorithm.
The algorithm consists of the following steps: (i) choosing a primary direction of movement, (ii) swimming in the selected direction, (iii) reorientation of bacteria with some probability.

We first consider the dynamics in between two consecutive reorientation events, i.e., steps (i) and (ii). 
Bacteria swim in a fluid medium and thus are subjected to thermal fluctuations. 
In addition there can be fluctuations of other origins, e.g. coming from molecular flagellar motors. 
In our model all these effects are adsorbed into the fluctuations of the cell's swimming direction.
In terms of the cell velocity this is the standard rotational diffusion (since the absolute value of the velocity remains constant) and thus 
the process can be parametrized with two angles of the spherical coordinate system, $\theta$ and $\varphi$. Coordinates of the cell are then obtained by integrating the velocity vector in time.

Langevin dynamics of the angles is described by a pair of stochastic equations,
\begin{equation}
	\label{eq:1}
	\dot{\theta} = D_r\cot\theta +\sqrt{2D_r}\xi_{\theta}(t),~~~
	\dot{\varphi} = \frac{\sqrt{2D_r}}{\sin\theta}\xi_{\varphi}(t),
\end{equation}
where $\xi_{\theta}(t)$ and $\xi_{\varphi}(t)$ are two independent standard Gaussian random variables of dispersion one,
and $D_r$ is the rotational diffusion coefficient.

These stochastic equations results in a  Fokker-Planck equation for the probability density  $P(\theta,\varphi,t)$:
\begin{equation}
	\label{eq:2}
	\frac{\partial P}{\partial t} = D_r \nabla P = 
	D_r \left[ \frac{1}{\sin \theta}\frac{\partial}{\partial \theta}(\sin\theta \frac{\partial}{\partial \theta}) + \frac{1}{\sin^2 \theta}\frac{\partial^2}{\partial \varphi^2}\right]P .
\end{equation}

More practically, if the time propagation step is short, the new direction of the velocity vector (with respect to the initial direction) is given
by $\phi \backsimeq \arccos{(1 - 2 D_r \Delta t)}$. A new vector should be chosen randomly on the corresponding  cone.

Next we consider a reorientation event, step (iii). 
In an environment with a chemoattractant gradient, the  
run intervals are no longer isotropic in space; they depend (statistically) on the direction of the motion. They change differently depending on whether the bacterium  
moves up or down the  gradient. 
The statistics (rate) of changes is determined,  according to the linear theory \cite{Block1983},   by Eq.~ (1).
The specific form of the memory kernel, Eq.~(2), allows to transform the original non-Markovian (non-local in time) dynamics 
into a local one by  extending the dimension of the space of dynamical variables. This technique is well-known in the field of 
stochastic processes driven by colored noise where it is  called  ``embedding''; see, e.g., Refs.~\cite{embeding1,embeding2}.

Avoiding technical details, we present the final results \cite{Celani2009}. The variation of the reorientation rate is given by
\begin{equation}
	\label{eq:5}
	\Delta(t) = \int_{-\infty}^{t} R(t-\tau) c(\tau) d\tau  = m_0(t) - \frac{m_1(t)}{2} - \frac{m_2(t)}{4},
\end{equation}
where three additional variables $m_i$ are determined form a system of coupled linear differential equations,
\begin{equation}
	\label{eq:4}
	\begin{cases}
		\dot{m_0} = -\lambda_0 m_0 + c(t), \\
		\dot{m_1} = -\lambda_0 m_1 + m_0, \\
		\dot{m_2} = -\lambda_0 m_2 + m_1.
	\end{cases}
\end{equation}
where 
$c(t) = c(\boldsymbol{r}(t))$ is the chemical concentration, measured at the location of the bacterium $\boldsymbol{r}(t)$ at the time $t$.

Now we are ready to describe the numerical algorithm. After every integration step, 
the bacterium re-orients itself in a new random direction with the probability $p = \min (1, \lambda(t) \Delta t)$, 
where $\Delta t$ is the length of single integration step,  or, alternatively,  it continues to move in the same direction with probability $1-p$.
Finally, the algorithm is a sequence of the following steps:

\begin{enumerate}
	\item Calculate reorientation frequency $\lambda(t)$ for the current state of the bacteria using linear chemotaxis theory. 
	For this, solve the system of equations (\ref{eq:4}), e.g., by using Euler's method with small step $\Delta t$. 
	Reorientation of bacteria occurs with probability $p = \min (1, \lambda(t) \Delta_r t)$ according to the pattern. Thereby, calculate the new bacteria speed vector.
	\item Calculate the new bacteria position at the time $\Delta t$ (by using the standard linear Euler scheme).
	\item For applying the rotational diffusion, calculate a 
	new speed vector, i.e. rotate it by an angle $\phi =\arccos{(1 - 2 D \Delta t)}$. 
	It gives a cone of directions, obtained by such rotation. We need to choose one of the directions on the cone, which we do by choosing on a unit circle (the base of the cone)
	a point following the uniform distribution. Thus we define  a new velocity vector. 
\end{enumerate}

\nolinenumbers


\begin{thebibliography}{40}
	
\bibitem{Whitman1998} 
Whitman WB, Coleman DC, Wiebe WJ. 
\newblock {{P}rokaryotes: {T}he unseen majority}. 
\newblock Proc Natl Acad Sci USA. 1998; 95: 6578–6583.

\bibitem{Kearns2010} 
Kearns DB.
\newblock {{A} field guide to bacterial swarming motility}. 
\newblock Nat Rev Microbiol. 2010; 8: 634–644.

\bibitem{OToole2000} 
O’Toole G, Kaplan HB, Kolter R.
\newblock {{B}iofilm formation as microbial development}.
\newblock Annu Rev Microbiol. 2010; 54: 49–79.

\bibitem{Stoodley2002}
Hall-Stoodley L, Costerton JW, Stoodley P.
\newblock {{B}acterial biofilms: from the natural environment to infectious diseases}.
\newblock Nat Rev Microbiol. 2002; 2: 95–108.

\bibitem{Jarrell2008}
Jarrell KF, McBride MJ.
\newblock {{T}he surprisingly diverse ways that prokaryotes move}.
\newblock Nat Rev Microbiol. 2008; 6 (6), 466-76.

\bibitem{Eisenbach2004} 
Eisenbach M.
\newblock {Chemotaxis}.
\newblock London: Imperial College Press, 1 edition; 2004.

\bibitem{Taktikos2013} 
Taktikos J, Stark H, Zaburdaev V.
\newblock {{H}ow the {M}otility {P}attern of {B}acteria {A}ffects {T}heir {D}ispersal and {C}hemotaxis}.
\newblock PLoS ONE. 2013; 8(12): e81936.

\bibitem{Berg1972} 
Berg HC, Brown DA.
\newblock {{C}hemotaxis in {E}scherichia coli analysed by three-dimensional tracking}.
\newblock Nature. 1972; 239: 500–504.

\bibitem{Turner2000} 
Turner L, Ryu WS, Berg HC.
\newblock {{R}eal-time imaging of fluorescent agellar filaments}.
\newblock J Bacteriol. 2000; 182: 2793–2801.

\bibitem{Theves2013}
Theves M, Taktikos J, Zaburdaev V, Stark H, Beta C.
\newblock {{A} bacterial swimmer with two alternating speeds of propagation}.
\newblock Biophys J. 2013; 105: 1915–1924.

\bibitem{Johansen2002}
Johansen JE, Pinhassi J, Blackburn N, Zweifel UL, Hagström Å.
\newblock {{V}ariability in motility characteristics among marine bacteria}.
\newblock Aquat Microb Ecol. 2002; 28(3): 229-237.

\bibitem{Locsei2007} 
Locsei JT.
\newblock {{P}ersistence of direction increases the drift velocity of run and tumble chemotaxis}.
\newblock J Math Biol. 2007; 55:41–60.

\bibitem{deGennes2004} 
de Gennes PG.
\newblock {{C}hemotaxis: the role of internal delays}.
\newblock Eur Biophys J. 2004; 33: 691–693.


\bibitem{Xie2011}
Xie L, Altindal T, Chattopadhyay S, Wu XL.
\newblock {{B}acterial flagellum as a propeller and as a rudder for efficient chemotaxis}. 
\newblock Proc Natl Acad Sci USA. 2011; 108: 2246–2251.

\bibitem{Son2013}
Son K, Guasto JS, Stocker R.
\newblock {{B}acteria can exploit a flagellar buckling instability to change direction}.
\newblock Nat Phys. 2013; 9: 494–498.

\bibitem{Berg2013}
Berg HC.
\newblock {{C}ell motility: {T}urning failure into function}.
\newblock Nat Phys. 2013; 9: 460–461.

\bibitem{Taute2015}
Taute KM, Gude S, Tans SJ, Shimizu TS.
\newblock {{H}igh-throughput 3D tracking of bacteria on a standard phase contrast microscope}.
\newblock Nat Comm. 2015; 6:8776.

\bibitem{Son2016}
Son K, Menolascina F, Stocker R.
\newblock {{S}peed-dependent chemotactic precision in marine bacteria}.
\newblock Proc Natl Acad Sci USA. 2016; 113: 8624–8629.

\bibitem{Theves2015}
Theves M, Taktikos J, Zaburdaev V, Stark H, Beta C.
\newblock {{R}andom walk patterns of a soil bacterium in open and confined environments}.
\newblock EPL. 2015; 109 (2), 28007.

\bibitem{Liu2014}
Liu B, Gulino M, Morse M, Tang JX, Powers TR, Breuer KS
\newblock {{H}elical motion of the cell body enhances {C}aulobacter crescentus motility}.
\newblock Proc Natl Acad Sci USA. 2014; 111: 11252–112566.

\bibitem{Altindal2011}
Altindal T, Xie L, Wu XL.
\newblock {{I}mplications of three-step swimming patterns in bacterial chemotaxis}.
\newblock Biophys J. 2011; 100: 32–41.

\bibitem{footnote1}
For possible affects of non-exponential behavior see \cite{Xie2011}.

\bibitem{Kafri2008}
Kafri Y, da Silveira RA.
\newblock {{S}teady-state chemotaxis in escherichia coli}.
\newblock Phys Rev Lett. 2008; 100: 238101.

\bibitem{Block1983}
Block SM, Segall JE, Berg HC.
\newblock {{A}daptation kinetics in bacterial chemotaxis}.
\newblock J Bacteriol. 1983; 154:312–323.

\bibitem{Celani2009}  
Celani A, Vergassola M.
\newblock {{B}acterial strategies for chemotaxis response}.
\newblock Proc Natl Acad Sci USA. 2009; 107: 1391-1396.

\bibitem{Xie2015}
Xie L, Altindal T, Wu XL.
\newblock {{A}n {E}lement of {D}eterminism in a {S}tochastic {F}lagellar {M}otor {S}witch}.
\newblock PLoS ONE. 2015; 10(11): e0141654.

\bibitem{XieLu2015}
Xie L, Lu C, Wu XL.
\newblock {Marine bacterial chemoresponse to a stepwise chemoattractant stimulus}.
\newblock Biophys J. 2015; 108: 766-74.

\bibitem{embeding1} 
Grabert H, Talkner P, H\"{a}nggi P.
\newblock {{M}icrodynamics and {T}ime-{E}volution of {M}acroscopic {N}on-{M}arkovian {S}ystems}.
\newblock Z Physik B. 1977; 26, 389.

\bibitem{embeding2} 
Kupferman R.
\newblock {{F}ractional kinetics in {K}ac-{Z}wanzig heat bath models}.
\newblock J Stat Phys. 2004; 114, 291.

\bibitem{Stocker2011}
Stocker R.
\newblock {{R}everse and flick: {H}ybrid locomotion in bacteria}. 
\newblock Proc Natl Acad Sci USA. 2011; 108: 2635–2636.

\bibitem{Bubendorfer2014} 
Bubendorfer S, Koltai M, Rossmann F, Sourjik V, Thormann KM.
\newblock {{S}econdary bacterial flagellar system improves bacterial spreading by increasing the directional persistence of swimming}.
\newblock Proc Natl Acad Sci USA. 2014; 111: 11485–11490.

\bibitem{Xie2014} 
Xie L, Wu XL.
\newblock {{B}acterial {M}otility {P}atterns {R}eveal {I}mportance of {E}xploitation over {E}xploration in {M}arine {M}icrohabitats. {P}art I: {T}heory}.
\newblock Biophys J. 2014; 107: 1712–1720.

\end{thebibliography}
\end{document}